\documentclass{IEEEtran}
\usepackage{cite}
\usepackage{amsmath,amssymb,amsfonts}
\usepackage{graphicx}
\usepackage{textcomp,nicefrac}
\usepackage{lineno}
\def\BibTeX{{\rm B\kern-.05em{\sc i\kern-.025em b}\kern-.08em
T\kern-.1667em\lower.7ex\hbox{E}\kern-.125emX}}

\usepackage[colorlinks=true, linkcolor=black, citecolor=black, urlcolor=black]{hyperref}

\newenvironment{keywords}{
	\vspace{1em}
	\textbf{\textit{Keywords ---}}\hspace{1em}
	\begin{itshape}
	}{
	\end{itshape}
}

\begin{document}

\title{Results of the LEGEND-200 experiment \\ in the search for $0\nu\beta\beta$ decay}
\author{C. Romo-Luque on behalf of the LEGEND Collaboration \\ \IEEEmembership{Los Alamos National Laboratory, Los Alamos, NM 87545, USA}
}

\maketitle

\begin{abstract}
The LEGEND experiment is looking for the extremely rare neutrinoless double beta ($0\nu\beta\beta$) decay of $^{76}$Ge using isotopically-enriched high-purity germanium (HPGe) detectors. The detection of this process would imply that the neutrino is a Majorana particle and the total lepton number would not be conserved, which could be related to the cosmological asymmetry between matter and antimatter through leptogenesis. The long-term goal of the collaboration is LEGEND-1000: a 1-ton detector array planned to run for 10 years, with a projected half-life sensitivity exceeding $10^{28}$ years, fully covering the inverted neutrino mass hierarchy. A first search for the  $0\nu\beta\beta$ decay has been carried out by LEGEND-200 building on the experience gained from GERDA and the MAJORANA DEMONSTRATOR. The experiment has been collecting physics data for a year at the Gran Sasso National Laboratory in Italy with 140 kg of HPGe detectors. With a total exposure of 61 kg yr, LEGEND-200 has achieved a background index of $5^{+3}_{-2}\times10^{-4}$ counts/(keV kg yr) in the $0\nu\beta\beta$ decay signal region from the highest performing detectors. After combining the results from GERDA, the MAJORANA Demonstrator and LEGEND-200, an exclusion sensitivity  $ > 2.8\times10^{26}$ yr has been obtained at 90\% confidence level for the $0\nu\beta\beta$ decay half-life, with no evidence for a signal. A new observed lower limit of $T^{0\nu}_{1/2} > 1.9\times10^{26}$ yr at 90\% confidence level has been established.

\end{abstract}


\section{Introduction}
\label{sec:introduction}

Neutrino oscillation experiments have established that neutrinos possess mass and that lepton flavor is not conserved—findings that imply physics beyond the Standard Model. At the same time, the observed imbalance between matter and antimatter in the Universe remains one of the most profound unsolved problems in cosmology. A process that could provide insight into these mysteries is neutrinoless double beta decay ($0\nu\beta\beta$). First proposed by Furry in 1939 \cite{Furry:1939qr}, this hypothetical nuclear transition would involve the emission of two electrons without accompanying neutrinos, and could occur if neutrinos are Majorana particles—that is, if they are their own antiparticles \cite{Majorana:1937vz}. The process would signify violation of lepton number conservation and offer a potential window into the origin of the matter-antimatter asymmetry, as well as unresolved questions about neutrino properties, such as their absolute mass and mass hierarchy.

In $0\nu\beta\beta$, the decay proceeds via the exchange of light Majorana neutrinos. Unlike standard two-neutrino double beta decay ($2\nu\beta\beta$), in which the emitted electrons share energy with neutrinos, the neutrinoless mode would result in the electrons carrying the full decay energy. This would manifest as a narrow peak in the summed electron energy spectrum, just at the endpoint of the broad spectrum produced by the standard process with neutrinos.

Current experimental limits place the half-life of $0\nu\beta\beta$ decay at greater than $10^{26}$ years. As a result, any experiment searching for this rare process must fulfill several stringent criteria: an extremely low background rate to maximize discovery sensitivity, excellent energy resolution to clearly identify the expected peak at the decay energy, large exposure (i.e., a high product of detector mass and measurement time) to increase the probability of observing a decay, and a high Q$_{\beta\beta}$ value to minimize interference from low-energy backgrounds.

\section{Germanium for $0\nu\beta\beta$ searches}

\begin{figure}[t]
	\centerline{\includegraphics[width=3in]{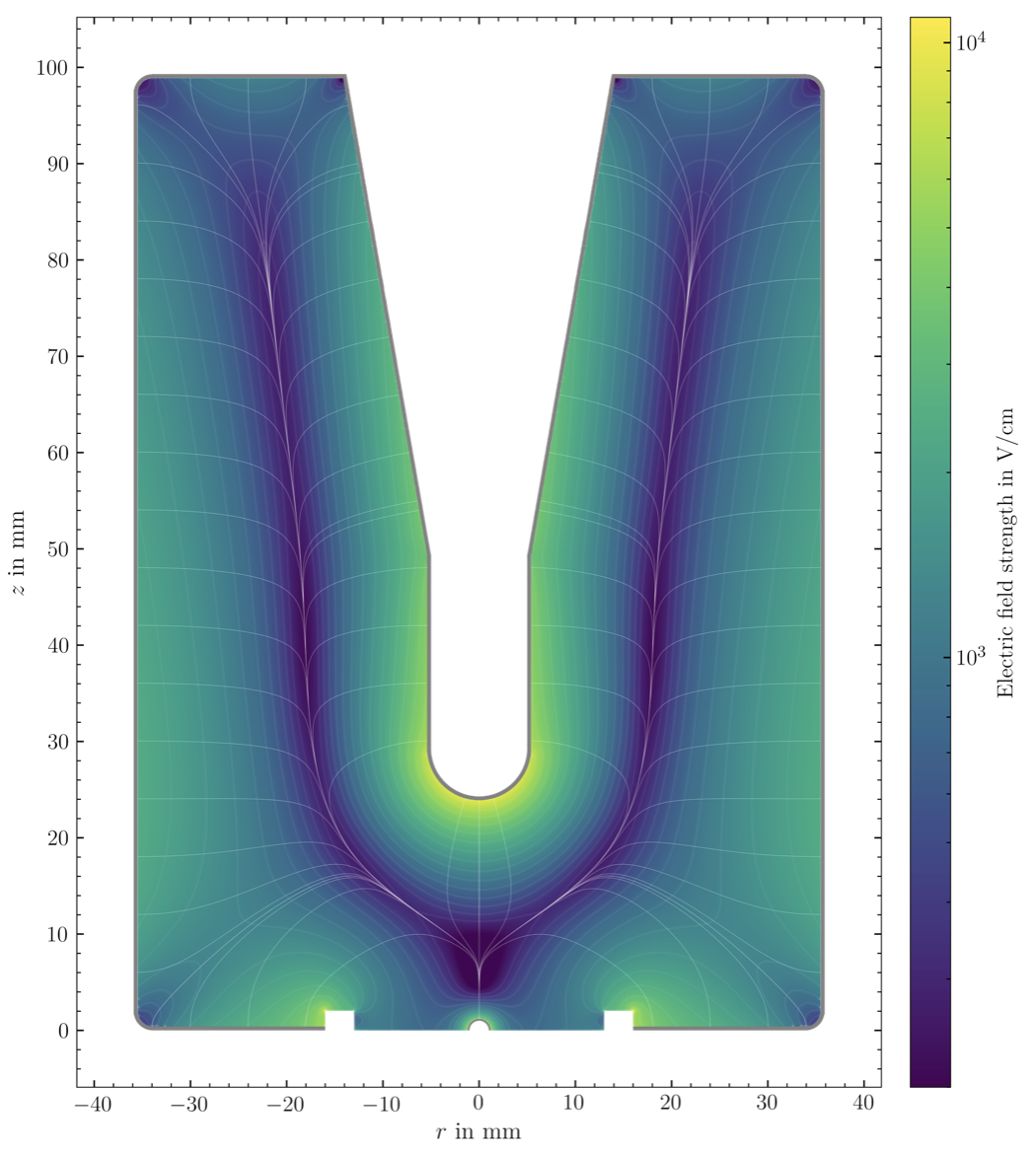}}
	\caption{Electric field lines of an inverted coaxial point contact (ICPC) detector. Contacts are shown in gray.}
	\label{fig1}
\end{figure}

There are 35 known isotopes that can undergo 2-neutrino double beta decay, of which 11 have been experimentally observed. Among them, $^{76}$Ge—decaying to $^{76}$Se—is one of the most well-suited candidates for $0\nu\beta\beta$ searches. High-purity germanium (HPGe) detectors can be fabricated using enriched $^{76}$Ge, offering several key advantages for $0\nu\beta\beta$ experiments. Germanium detectors achieve the best energy resolution of any $0\nu\beta\beta$ experiment, reaching approximately 0.1\% FWHM at Q$_{\beta\beta}$, which allows precise discrimination between signal and background. Additionally, they offer detailed information on the event topology, enabling improved identification of true decay events and further suppression of background noise. These detectors can be isotopically enriched beyond 90\% in $^{76}$Ge resulting in a very low intrinsic background. In addition, in HPGe detectors the source of the decay and the active volume are the same, which leads to high detection efficiency with minimal signal loss and do not require self-shielding. Germanium detectors have undetectably low $^{232}$Th- and $^{238}$U-chain internal contamination and no known background source produces a peak near Q$_{\beta\beta}$. Fig. \ref{fig1} shows an HPGe inverted coaxial detector with the electric field lines.

Germanium has been studied for over 50 years, with notable efforts in the last decade led by the GERDA and MAJORANA DEMONSTRATOR (MJD) collaborations. Thanks to the complementarity of the LAr scintillation light detection and the event topology reconstruction of Ge detectors, GERDA achieved the world-leading background index of $5.2^{+1.6}_{-1.3} \times 10^{-4}$ cts/(keV kg yr) around $Q_{\beta\beta}$ and set one of the most stringent limits on the $0\nu\beta\beta$ half-life ($> 1.8 \times 10^{26}$ yr at 90\% CL) \cite{GERDA:2020xhi}. MJD developed low noise electronics allowing to achieve a record energy resolution for large-scale $0\nu\beta\beta$ decay experiments of $2.52 \pm 0.08$ keV (0.124\%) FWHM \cite{Majorana:2022udl}.

In HPGe detectors, particle interactions within the detector generate electron-hole pairs. The applied electric field drives electrons toward the n-type contact and holes toward the p-type contact, inducing signal in the p-type contact. The total charge is proportional to the energy of the event and the resulting charge and current drift patterns differ between signal and background events, enabling more effective background discrimination.

\section{The LEGEND project}

The Large Enriched Germanium Experiment for Neutrinoless $\beta\beta$ Decay (LEGEND) \cite{LEGEND:2021bnm} is an international collaboration established in 2016 to unite the technical expertise and leadership from both the MJD and GERDA collaborations, while also incorporating new members to strengthen core capabilities. Today LEGEND comprises nearly 300 collaborators from 60 institutions across 12 countries.

The LEGEND project is designed to carry out a two-phase experimental campaign with the ultimate goal of $0\nu\beta\beta$ decay discovery potential for half-lives exceeding $10^{28}$ years. The first phase, LEGEND-200, is currently running in an upgrade of the GERDA infrastructure at the Laboratori Nazionali del Gran Sasso (LNGS) in Italy. LEGEND-200 is expected to operate with 200 kg of enriched germanium detectors immersed in liquid argon (LAr) over a period of 5 years, achieving a total exposure of 1 ton yr. The target background index is $2 \times 10^{-4}$ counts/(keV kg yr) and the projected discovery sensitivity for the $0\nu\beta\beta$ half-life is around $10^{27}$ yr at $3\sigma$ confidence level. The second phase, LEGEND-1000, will use 1 ton of enriched HPGe detectors and will operate for 10 years at Hall C at LNGS. By conducting a search with near-zero background, it aims to cover the inverted-ordering neutrino mass scale region, with a projected $0\nu\beta\beta$ half-life discovery sensitivity $> 1.3\times10^{28}$ yr.

\section{LEGEND-200}

\begin{figure}[t]
	\centerline{\includegraphics[width=3.5in]{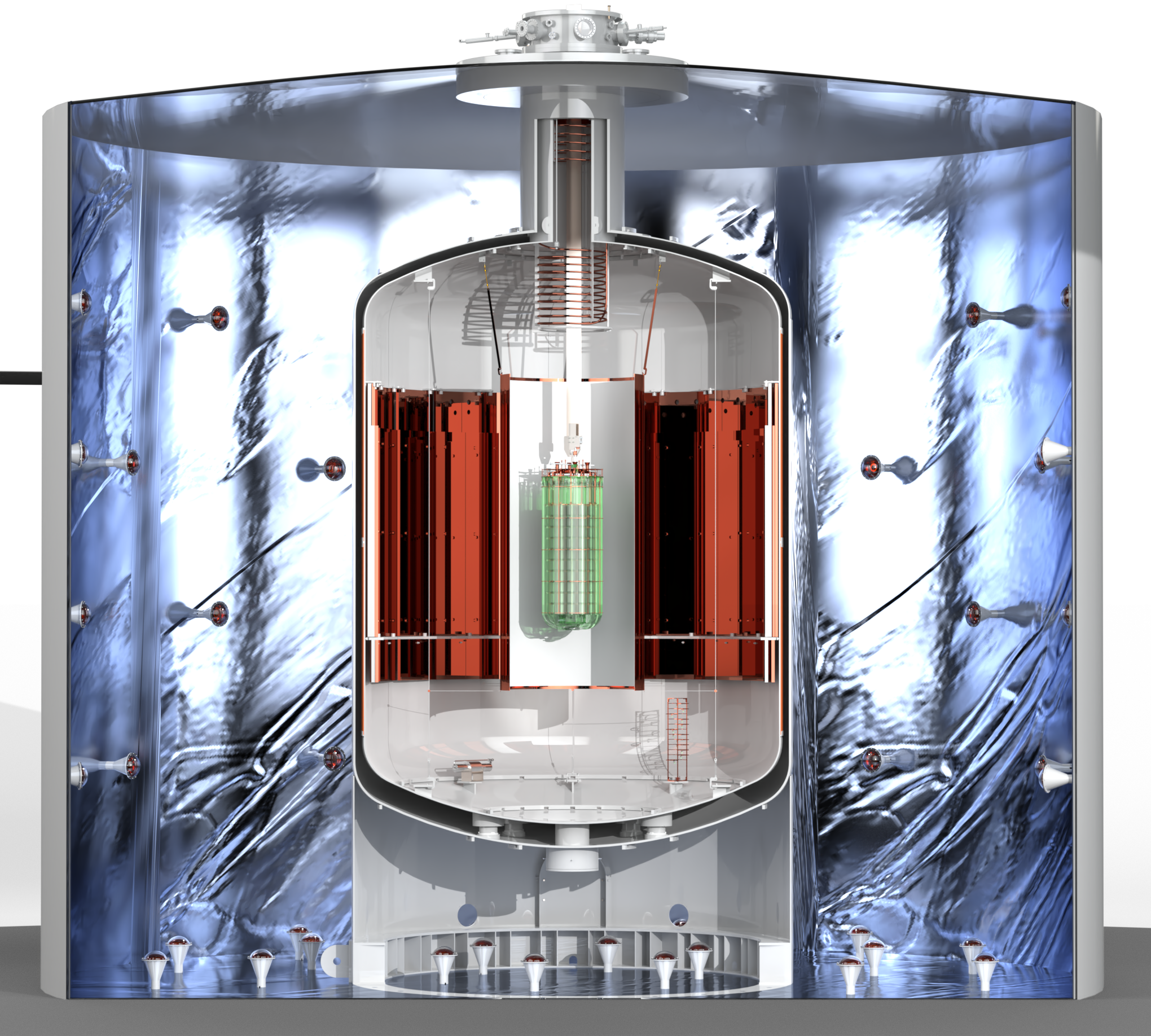}}
	\begin{picture}(0,0)
		\put(125,120){\makebox(0,0)[c]{\large\textbf{a}}}
		\put(118,132){\makebox(0,0)[c]{\large\textbf{b}}}
		\put(90,165){\makebox(0,0)[c]{\large\textbf{c}}}
		\put(70,180){\makebox(0,0)[c]{\large\textbf{d}}}
	\end{picture}
	\caption{Cross-sectional rendering of the LEGEND-200 experiment, showing (a) the arrangement of HPGe detectors, (b) the wavelength-shifting fibers coupled to SiPMs, (c) the liquid argon cryostat, and (d) the surrounding ultra-pure water tank.}
	\label{l200cryowater}
\end{figure}
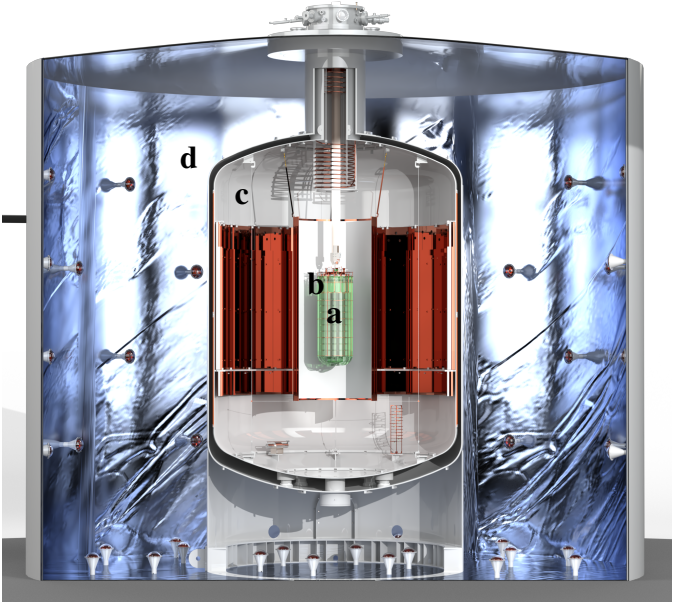

LEGEND-200 \cite{LEGEND:2025jwu}, shown in Fig. \ref{l200cryowater}, consists of a core of individual HPGe detectors enriched in $^{76}$Ge up to $86\% - 92\%$ and arranged in vertical strings. The detectors are mounted on scintillating polyethylene naphthalate plates using copper rods that are made from underground-electroformed Cu, with electrical isolation provided via plastic insulators. The detector arrays are immersed in a 64 m${^3}$ LAr bath, where the scintillation light from interacting particles in the argon is detected via two barrels of wavelength-shifting polystyrene fibers. One fiber barrel is positioned inside the circle formed by the detector strings and comprises 18 readout channels, while the outer barrel surrounds the array and contains 40 channels. The fibers are coated with TPB and coupled to silicon photomultiplier (SiPM) photodetectors, which provide single photo-electron detection capabilities. The granular nature of the Ge detector array and its immersion within an active LAr scintillating volume offers additional strong discrimination between $0\nu\beta\beta$ decay signal events, which are isolated to a single Ge detector, and background-generating events in multiple coincident detectors.

An ultra-pure water tank instrumented with 63 photomultiplier tubes (PMTs) surrounds the argon cryostat and serves as a passive shielding from muon-induced neutrons preventing them from reaching sensitive detectors and as veto for through-going muons based on Cherenkov light. The inner sufaces are covered with a reflective, wavelength-shifting film that enhances the light yield and converts the scintillation light to 400 nm, matching the sensitivity range of the PMTs.

In October 2022, 142 kg of enriched germanium detectors were installed at LNGS. The underground location of the laboratory, with approximately 3500 m water equivalent of overburden, provides effective shielding from external radiation and cosmic rays, significantly reducing backgrounds. A total of 101 Ge detectors were deployed, with four different geometries: P-type Point Contact (PPC) from MJD, Broad Energy Germanium (BEGe) plus semicoaxial (Coax) from GERDA and the novel Inverted-Coaxial Point-Contact (ICPC) specifically designed for LEGEND. Detector masses ranged from 0.7 kg in the BEGe units to up to 4 kg in some of the larger ICPC detectors.

\section{LEGEND-200 results}

The first physics data-taking period of LEGEND-200 started in March 2023 and concluded in February 2024, with a total exposure of 61 kg yr selected for the final $0\nu\beta\beta$ decay analysis. After a commissioning phase, a blinding strategy was implemented: all events with an energy between $\pm$ 25 keV around Q$_{\beta\beta}$ were withheld to prevent bias during analysis.

\begin{figure}[t]
	\centerline{\includegraphics[width=3.5in]{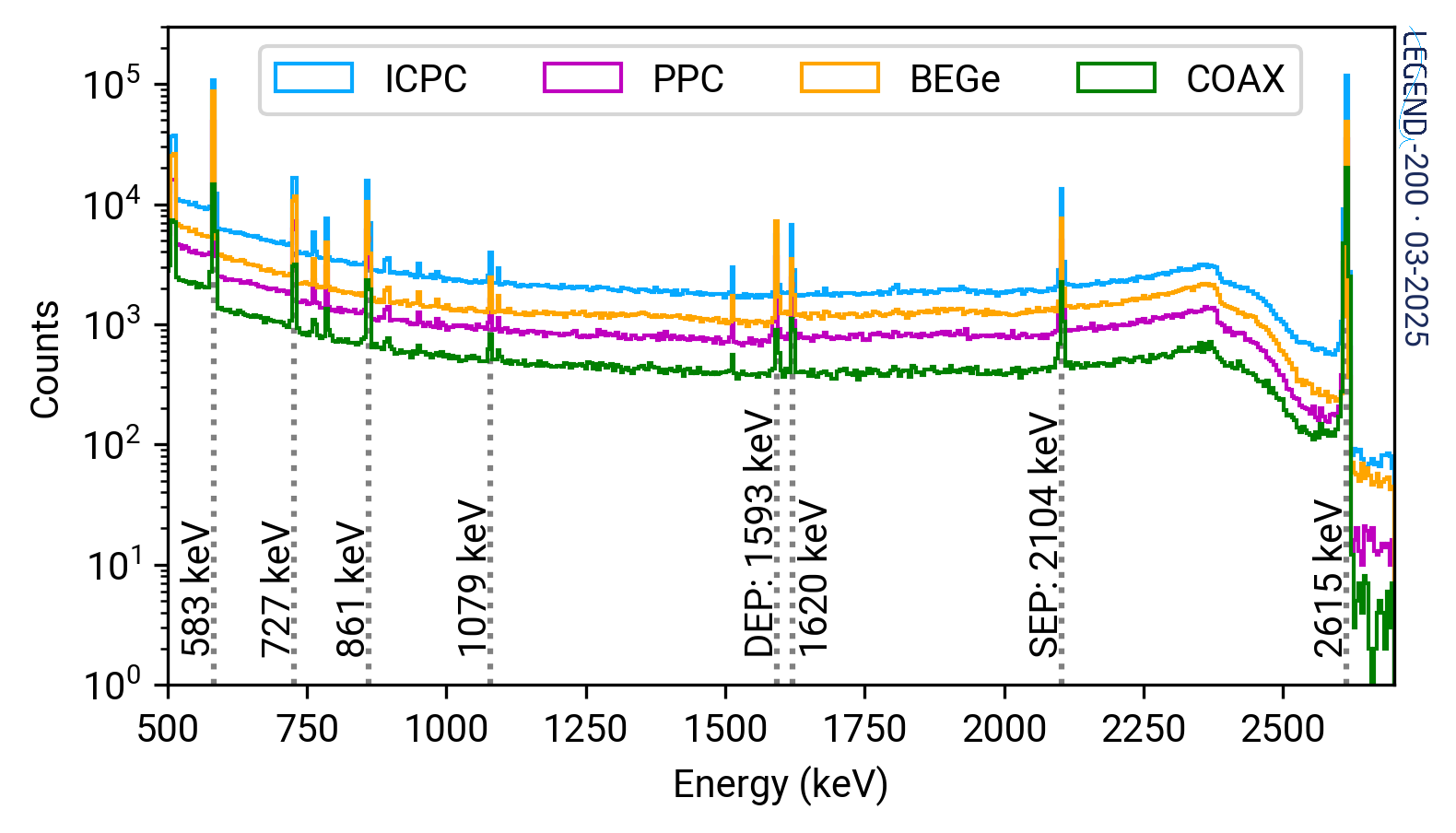}}
	\centerline{\includegraphics[width=3.4in]{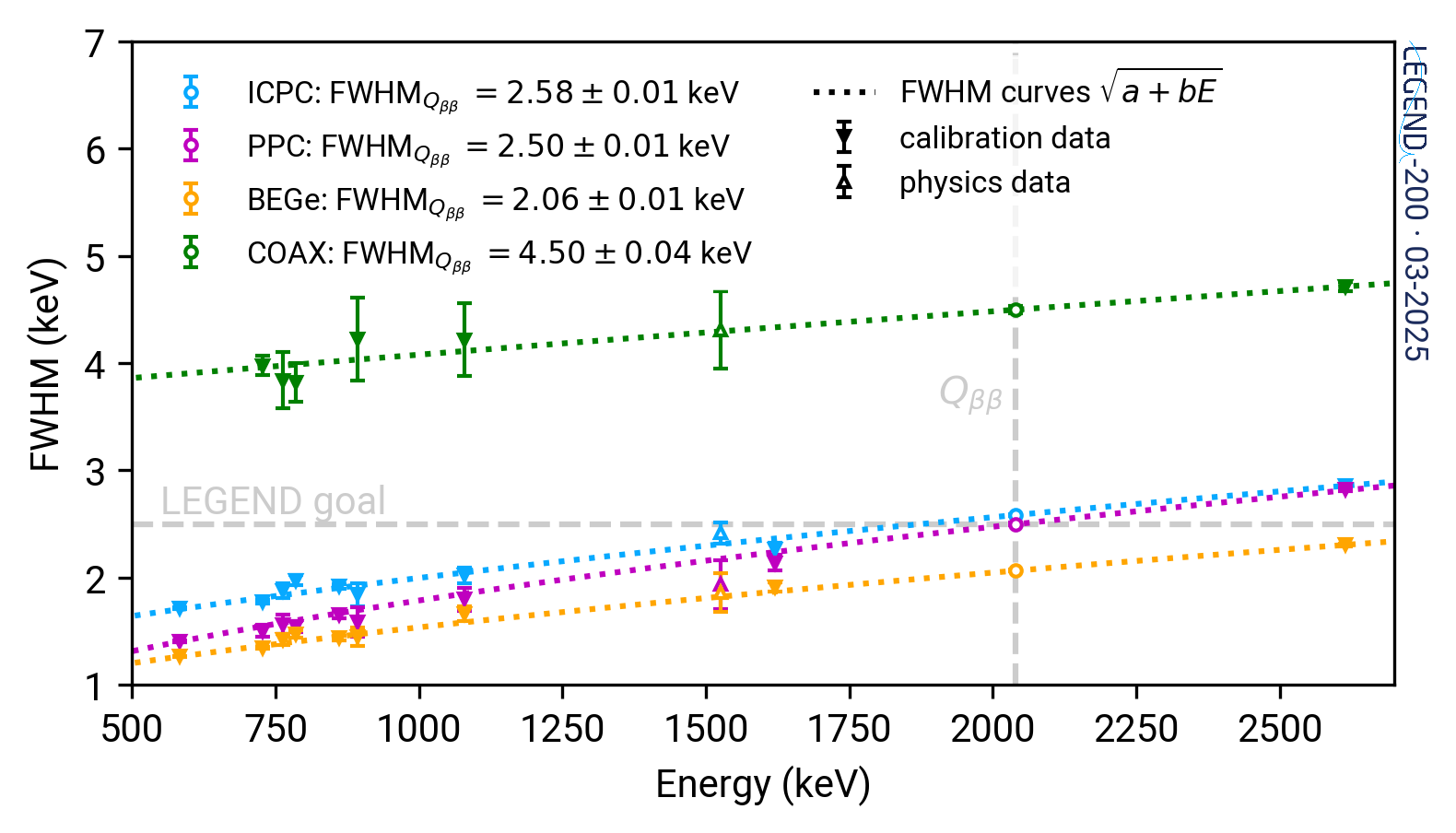}}
	\caption{Top: energy spectrum from a $^{228}$Th calibration run for the four different detector types. Bottom: FWHM energy resolution curves (dotted line) derived from calibration data as a function of energy for the four different detector types. For completeness, energy resolutions from physics data at the $^{42}$K line (1525 keV) are included. The empty circle indicates the extrapolated resolution at Q$_{\beta\beta}$. The LEGEND-200 energy resolution goal of 2.5 keV at Q$_{\beta\beta}$ is shown in the dashed gray line.}
	\label{eres}
\end{figure}

Quality cuts are used in the analysis to exclude events that do not match typical energy depositions in the HPGe array and events with energy depositions in multiple HPGe detectors (multiplicity cut) are discarded. Calibrations are performed weekly using twelve $^{228}$Th sources of around 5 kBq each \cite{Baudis:2022lcu} to assess the energy scale, resolution and pulse shape discrimination (PSD) performance of the HPGe detectors. Between calibrations, physics runs are taken. Data are split into partitions, that is, different stable periods for each HPGe. Within each partition, the energy resolution, the energy bias and the PSD parameters remain without significant variations for a certain number of runs. The sensitivity to $0\nu\beta\beta$ decay critically depends on both energy scale stability and superior energy resolution. Fig. \ref{eres} shows the energy spectrum from a calibration run (top) and the energy resolution curves (bottom) for the different detector types. All detectors except the Coax (which are not planned for future configurations) meet the LEGEND requirements for energy resolution, with stable values of 0.1\% FWHM at Q$_{\beta\beta}$ throughout the entire data taking.

Muon signals recorded by the PMTs in the water tank are processed offline, and events coincident with HPGe triggers within a $\pm1.75$ $\mu$s window are excluded from the dataset.

\begin{figure*}[t]
		\centerline{\includegraphics[width=\textwidth]{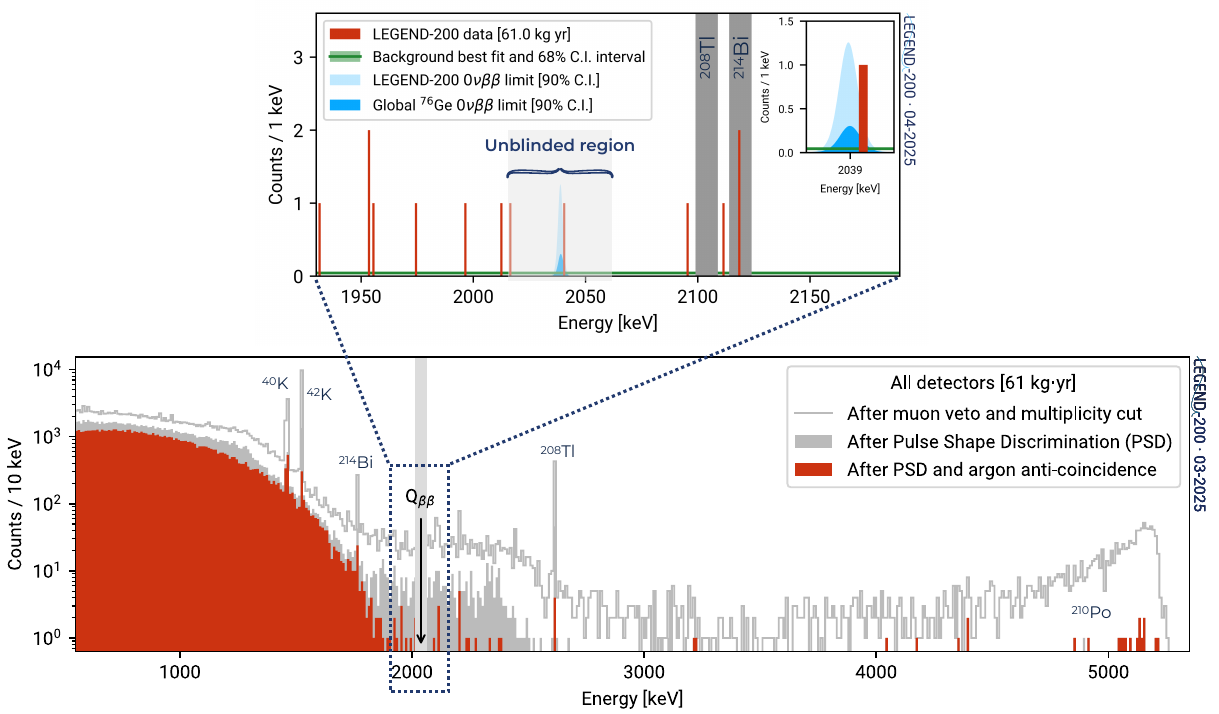}}
		\caption{Energy spectrum of LEGEND-200 from its first year of physics data collection, with a total exposure of 61 kg yr. An energy window of $\pm$25 keV around $Q_{\beta\beta}$ was blinded. The gray histogram represents events that passed the quality, muon veto and multiplicity cuts, with the major background sources labeled. The filled gray histogram includes events that also passed pulse shape discrimination, resulting in a significant suppression of surface-origin $\alpha$ and $\beta$ events. The subset of events that additionally passed the anti-coincidence liquid argon (LAr) cut is shown in red, where remaining Compton events and those from the $^{232}$Th decay chain are further strongly suppressed. The inset above shows the unblinded analysis window with the events surviving all cuts. The known gamma lines ($\pm5$ keV) are indicated by gray bands, which are excluded from the statistical analysis.}
		\label{l200_phy_spec_psd_and_all}
	\end{figure*}

The gray line in Fig. \ref{l200_phy_spec_psd_and_all} shows the energy spectrum of the acquired data after quality, HPGe multiplicity and muon veto cuts. Below $Q_{\beta\beta}$ the continuous distribution of $2\nu\beta\beta$ decay events dominates the spectrum. Gamma rays from the decays of $^{40}$Ar, $^{238}$U and $^{232}$Th in structural materials, as well as from the decay of $^{42}$K (progeny of $^{42}$Ar) in liquid argon, contribute to the gamma lines and continuum observed below 3 MeV. At higher energies, the spectrum is dominated by energy-degraded $\alpha$ particles from the $^{238}$U chain. In addition, $\beta$ particles from the decay of $^{42}$K, which has a $Q_{\beta}$ value of 3.53 keV, can reach the detector and penetrate into the active volume. While the spectrum shows all anticipated features, comparison with radioassay-based background predictions indicates an elevated background level, including near $Q_{\beta\beta}$.

Two PSD techniques are employed to identify signal-like events based on the rising edge of the HPGe waveforms. The first is the \textit{A/E} classifier, which measures the ratio of the maximum current amplitude (\textit{A}) to the total energy (\textit{E}) of the event. This method is effective for rejecting multi-site events, which typically exhibit lower \textit{A/E} values compared to single-site events, and $\alpha$-particle interactions, characterized by higher \textit{A/E} values. The second PSD classifier, late charge (LQ), is defined as the area above the last 20\% of the charge signal normalized by the energy. Events near the electrodes have high LQ values.

Scintillation light registered in the liquid argon veto in coincidence with signals from the HPGe detectors is used to suppress background. The coincident window is defined as [-1, 5] $\mu$s relative to the onset of the HPGe rising edge. Events are excluded from the analysis if the total amplitude summed across all SiPM channels exceeds four photoelectrons, or if more than four channels register signals above threshold. 
The subsets of events passing the PSD cut and the combined PSD plus LAr veto cuts are shown in Fig. \ref{l200_phy_spec_psd_and_all} as the gray-filled and red histograms, respectively. After PSD, there is strong suppression of events originating in the surfaces, including $\alpha$ events from the $^{238}$U decay chain and $\beta$ events from $^{42}$K. Compton scatters from the 2615 keV gamma ray of $^{208}$Tl remaining after PSD are effectively removed by the LAr veto filtering selection. The resulting energy spectrum after all analysis cuts consists of a $2\nu\beta\beta$ distribution at lower energies, with only a few events surviving near $Q_{\beta\beta}$.

The inset of Fig. \ref{l200_phy_spec_psd_and_all} shows the energy spectrum after all analysis cuts in the range [1930-2190] keV, used for the statistical analysis and calculation of the background index. The gamma lines from $^{208}$Tl (2104 keV) and $^{214}$Bi (2119 keV) are excluded from the analysis with a window of $\pm$5 keV around each line. Two separate unblindings were performed for distinct subsets of data. The first, with an exposure of 48.3 kg yr, corresponds to the highest-performing HPGe detectors: BEGe, PPC and ICPC produced by Mirion. The second includes data from the Coax and ICPC detectors with abnormally wide p+ pads, with an exposure of 12.7 kg yr. As shown in the inset of Fig. \ref{l200_phy_spec_psd_and_all}, a total of eleven events survive all cuts—seven from the first subset and four from the second.

The results obtained with the frequentist analysis show no evidence of signal, setting a lower limit on the $0\nu\beta\beta$ decay half-life $> 0.5 \times 10^{26}$ yr at 90\% confidence level (C.L.). The background index is determined to be $5^{+3}_{-2}\times10^{-4}$ counts/(keV kg yr) and $13^{+8}_{-5}\times10^{-4}$ counts/(keV kg yr) for the first and second datasets, respectively. The elevated background index observed in the second dataset is attributed to the Coax detectors, which are not included in the plans for future phases of the experiment. The frequentist fit joining the results with GERDA (127.2 kg yr) and MJD (64.5 kg yr) data yields a combined lower limit on the $0\nu\beta\beta$ decay half-life of $1.9 \times 10^{26}$ yr (90\% C.L.), compatible with the Bayesian result \cite{LEGEND:2025jwu}. The combined exclusion sensitivity obtained is $> 2.8 \times 10^{26}$ yr, the best achieved among $0\nu\beta\beta$ decay searches to date.

After the initial year of physics data taking, a dedicated background characterization phase was conducted to investigate the sources of the observed background, which exceeded expectations. With the planned deployment of additional large-mass ICPC detectors, data acquisition will resume using an enhanced detector array.

\section{LEGEND-1000}

\begin{figure}[t]
	\centerline{\includegraphics[width=3.5in]{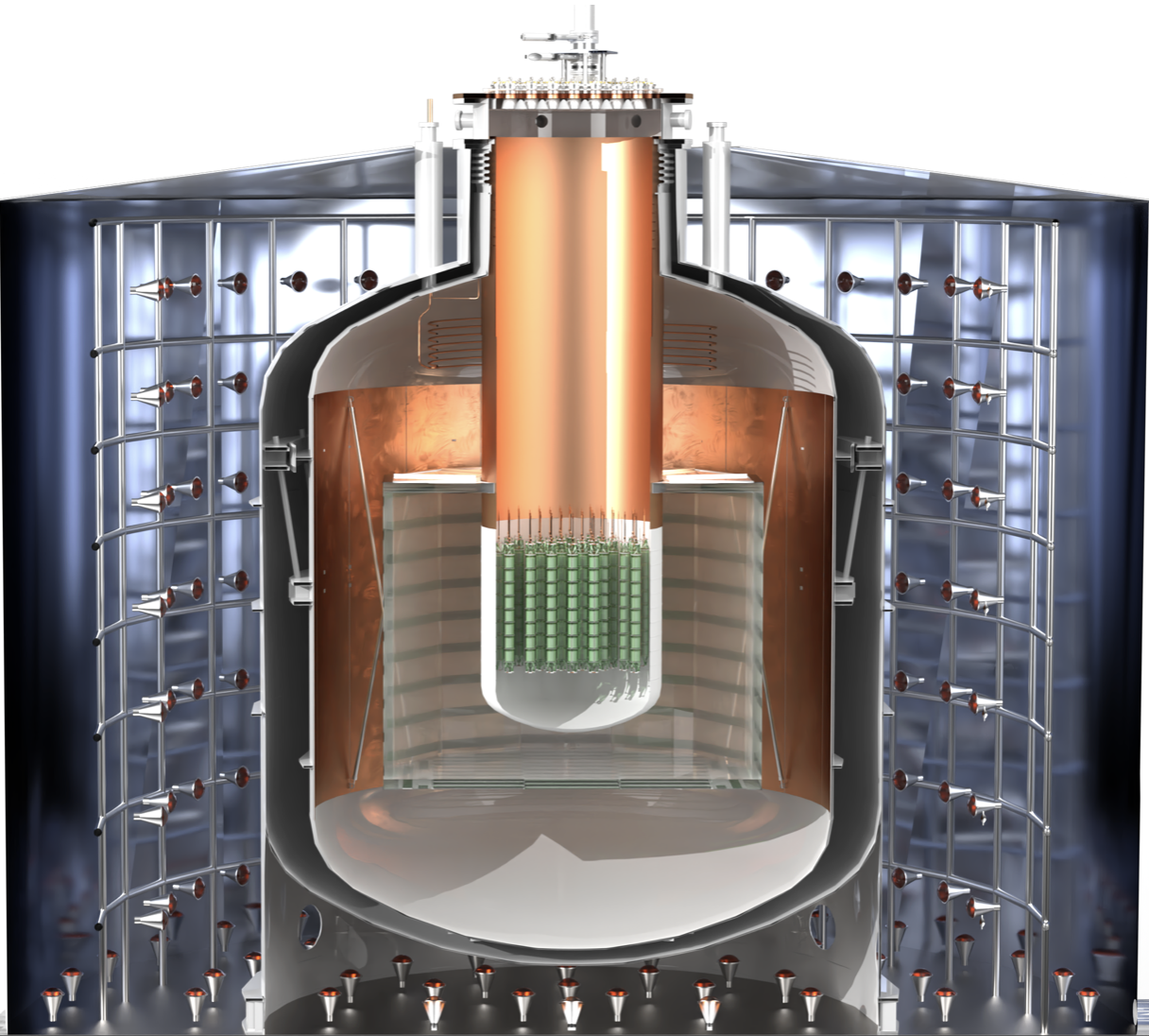}}
	\caption{Schematic design of the LEGEND-1000 experiment, featuring arrays of HPGe detectors at the center, immmersed in underground-sourced liquid argon. The neutron moderator surrounds the cryostat and is housed within atmospheric-sourced liquid argon, with the entire system enclosed in an ultra-pure water tank.}
	\label{l1000cryowater}
\end{figure}

LEGEND-1000 is planned to begin operation in the early 2030s, with 1 tonne of high-purity p-type, ICPC Ge semiconductor detectors, enriched above 90\% in $^{76}$Ge. These large detectors have demonstrated low background levels, excellent pulse shape discrimination and energy performance, reaching energy resolutions of 0.12\% FWHM (0.05\% $\sigma$) at $Q_{\beta\beta}$.

An additional 20-fold reduction in background is anticipated for LEGEND-1000 relative to LEGEND-200, achieved through the use of underground-sourced argon—reduced in the $^{42}$Ar isotope— newly developed low-radioactivity electronics and cabling, and the exclusive deployment of ICPC detectors.

The experiment, shown in Fig. \ref{l1000cryowater}, will be installed in Hall C at LNGS. It will re-deploy 130 kg of the ICPC detectors currently used in LEGEND-200, along with 870 kg of newly produced detectors. Approximately 340 ICPC detectors each with an average mass of 3.0 kg will be arranged in 42 strings to enable independent operation and phased commissioning. The strings will be supported by ultra-clean materials and read out using ultra-low-background application-specific itegrated circuit electronics. The detector arrays will be immersed in underground LAr and enclosed within an electroformed copper reentrant tube. The reentrant tube will be surrounded by conventional atmospheric-sourced LAr, contained within a vacuum-insulated cryostat. The argon volumes will be instrumented with a curtain of wavelength-shifting fibers and read out by SiPMs. The cryostat will be housed in a water tank providing additional shielding and instrumented with PMTs. A neutron moderator made of a hydrogen-rich polymer will be installed outside the underground argon volume to effectively thermalize cosmogenic neutrons, thereby reducing the probability of neutron capture on $^{76}$Ge. It will be instrumented with light guides read out by SiPMs to tag neutron captures on $^{40}$Ar, which produce a $\gamma$-ray cascade with a total energy of up to 6 MeV.

The Collaboration aims to construct the LEGEND-1000 experiment with the objective of probing $^{76}$Ge $0\nu\beta\beta$ decay half-lives beyond $10^{28}$ yr with a 3$\sigma$ discovery sensitivity, covering the full inverted neutrino mass hierarchy. LEGEND-1000 is optimized to operate in a quasi-background-free regime and is expected to enable an unambiguous discovery of $0\nu\beta\beta$ decay with a few detected counts at $Q_{\beta\beta}$.

\section{Conclusions}

The results from the first year of LEGEND-200 physics data taking, corresponding to a total exposure of 61 kg yr, have been presented. The achieved background levels, $5^{+3}_{-2}\times10^{-4}$ counts/(keV kg yr), while higher than predicted, are the lowest reported by any $0\nu\beta\beta$ experiment to date. No evidence of a $0\nu\beta\beta$ signal has been observed, and a lower limit on the half-life of $T_{1/2}^{0\nu} \geq 0.5 \times 10^{26}$ yr at 90\% C.L. has been set based on a frequestist analysis. A combined fit of data from GERDA, the MAJORANA Demonstrator and LEGEND sets a lower limit on the $^{76}$Ge $0\nu\beta\beta$ half-life at $T_{1/2}^{0\nu} \geq 1.9 \times 10^{26}$ yr (90\% C.L.). Data acquisition will resume with an enhanced detector array after deploying additional large-mass ICPC detectors and further background reduction by refining surface treatments of nearby components. LEGEND ultimately aims to operate up to 1 tonne of $^{76}$Ge-enriched HPGe detectors within the future LEGEND-1000. The program is designed to reach discovery sensitivities to $0\nu\beta\beta$ decay half-lives beyond $10^{28}$ years, thereby probing the inverted neutrino mass ordering.

\section*{Acknowledgment}

This work is supported by the U.S. DOE and the NSF, the LANL, ORNL and LBNL LDRD programs; the European ERC and Horizon programs; the German DFG, BMBF, and MPG; the Italian INFN; the Polish NCN and MNiSW; the Czech MEYS; the Slovak RDA; the Swiss SNF; the UK STFC; the Canadian NSERC and CFI; the LNGS and SURF facilities.


\begin{thebibliography}{00}
	
	\bibitem{Furry:1939qr}
	W.~H.~Furry, ``On transition probabilities in double beta-disintegration,''
	\href{https://doi.org/10.1103/PhysRev.56.1184}{Phys. Rev. \textbf{56}, 1184-1193 (1939).}
	
	\bibitem{Majorana:1937vz}
	E.~Majorana, ``Teoria simmetrica dell{\textquoteright}elettrone e del positrone,''
	\href{https://doi.org/10.1007/BF02961314}{Nuovo Cim. \textbf{14}, 171-184 (1937).}

	\bibitem{Majorana:2022udl}
	J.~Arnquist \textit{et al.} (Majorana), ``Final Result of the Majorana Demonstrator’s Search for Neutrinoless Double-$\beta$ Decay in $^{76}$Ge'', 
	\href{https://doi.org/10.1103/PhysRevLett.130.062501}{Phys. Rev. Lett. \textbf{130}, 062501 (2023)}.
	
	\bibitem{GERDA:2020xhi}
	M.~Agostini \textit{et al.} (Gerda), ``Final Results of GERDA on the Search for Neutrinoless Double-$\beta$ Decay,''
	\href{https://doi.org/10.1103/PhysRevLett.125.252502}{Phys. Rev. Lett. \textbf{125}, no.25, 252502 (2020).}

	\bibitem{LEGEND:2021bnm}
	N.~Abgrall \textit{et al.} [LEGEND],
	``The Large Enriched Germanium Experiment for Neutrinoless $\beta\beta$ Decay: LEGEND-1000 Preconceptual Design Report,''
	[arXiv:2107.11462 [physics.ins-det]].
	
	\bibitem{LEGEND:2025jwu}
	H.~Acharya \textit{et al.} [LEGEND], ``First Results on the Search for Lepton Number Violating Neutrinoless Double Beta Decay with the LEGEND-200 Experiment,''
	[arXiv:2505.10440 [hep-ex]].
	
	\bibitem{Baudis:2022lcu}
	L.~Baudis, G.~Benato, E.~M.~Bond, P.~J.~Chiu, S.~R.~Elliott, R.~Massarczyk, S.~J.~Meijer and Y.~M{\"u}ller, ``Calibration sources for the LEGEND-200 experiment,''
	\href{https://doi.org/10.1088/1748-0221/18/02/P02001}{JINST \textbf{18}, no.02, P02001 (2023).}
	
	
\end{thebibliography}
\end{document}